\documentclass{elsart}
\usepackage{amssymb}
\usepackage{amsfonts}
\usepackage{amsmath}

\input{tcilatex}
\begin{document}

\begin{frontmatter}%

\title{Sorption Isotherms and Probability Theory of Complex Systems}%

\author{F.Brouers}%

\address{Institute of Chemical Engineering,University of Liège, Belgium} fbrouers@ulg.ac.be%

\begin{abstract}
We show that most of the empirical or semi-empirical isotherms proposed to
extend the Langmuir formula to sorption (adsorption, chimisorption and
biosorption) on heterogeneous surfaces in the gaseous and liquid phase
belong to the family and subfamily of the $Burr_{XII}$ cumulative
distribution functions. As a consequence they obey relatively simple
differential equations which describe birth and death phenomena resulting
from mesoscopic and microscopic physicochemical processes. Using the
probability theory, it is thus possible to give a physical meaning to their
empirical coefficients, to calculate well defined quantities and to compare
the results obtained from different isotherms. Another interesting
consequence of this finding is that it is possible to relate the shape of
the isotherm to the distribution of sorption energies which we have
calculated for each isotherm.\ In particular, we show that the energy
distribution corresponding to the Brouers-Sotolongo ($BS$) isotherm \cite%
{Bro05b} is the Gumbel extreme value distribution \ Finally we propose a
generalized $GBS$ isotherm, calculate its relevant statistical properties
and recover all the previous results by giving well defined values to its
coefficients. In the course of the discussion we make contact with the
Tsallis nonextensive theory \cite{Tsa09} and the noninteger order reaction
and fractal kinetics theory \cite{Bro06}. In the spirits of the present and
previous publications, we propose an alternative formula to include
fractality in the Michealis-Menten\ enzyme catalysis theory. Finally we
suggest that the stochastic cluster model introduced by K.Weron \cite{Wer02}
to account for the universal character of relaxation in disordered systems
should be relevant for other phenomena in particular for heterogeneous
sorption. 
\end{abstract}%


\end{frontmatter}%

\section{Introduction}

Every year hundreds or more papers are devoted to the analysis of sorption
(physical adsorption, chemi- and bio-sorption) of gas or solutions on a
variety of substrates \cite{Chi08}. Among them, a great number are concerned
with the decontamination of air, water and soil. One of the typical
procedures is a comparison of the data with empirical isotherm formulas
which in the course of time have been proposed by scientists working in the
field to generalize the original Langmuir isotherm to heterogeneous surfaces
and to sorption in solutions. Most of these formulas are empirical and bring
little information on the physicochemical processes responsible for the
particular shape of the isotherm curves. The evolution of the empirical
parameters with external factors are recorded but there is no precise
correlations between the variations of the parameters belonging to different
isotherms. It appears that some order should be introduced in that field in
order to propose well documented practical and useful applications.

In this paper which is a contribution to that effort, we want to emphasize
that since some of these isotherms appear to be genuine cumulative
probability distributions, they should be favoured, formulated in the
language of the theory of probability and might bring more quantitative and
more structured information making advantage of their mathematical
properties. The probability theory of complex systems has made considerable
progress these last years and one can expect that its introduction in the
field of sorption could be of great help.

\section{Sorption on heterogeneous surfaces}

A few years ago we published a paper \cite{Bro05b} actualizing the efforts
initiated by Langmuir, Zeldowitsch and followers eighty years ago to
incorporate in the classical Langmuir adsorption isotherm theory, the
heterogeneous nature of the substrate, the $N$-body interactions and the
nonequilibrium state of the sorbate. One important conclusion of this study
was that the most important ingredient playing a role in designing the shape
of the isotherm is the sorption energy distribution which itself is a
reflection of the disordered and complex nature of the phenomenon.\ In our
work, we insisted on the fact that it would be useful to rewrite the theory
in the framework of the theory of probability. Moreover we reminded that it
is an asymmetric birth and death (sorption-desorption) process and a rare
event dominated problem due to the very nature of the sorption mechanism,
the more active sites being the first to be occupied.\ We pointed out that
these characteristics should be taken into account in the theory. We showed
that to account for the power law Freundlich isotherm, one has to assume a L%
\'{e}vy heavy tail behavior for the temperature dependent Langmuir parameter.

\ As a consequence of this study we proposed to use an isotherm using a
Weibull distribution known as $Brouers-Sotolongo$ ($BS$) isotherm which
since has been used among others in sorption on porous/nonporous surface
interface \cite{Nci08}, magnetic nano-particles \cite{Mad12}, activated
carbon produced from natural products \cite{Var11,Alt09,Hej12}, algae, \cite%
{Nci09,Alt12}, soils \cite{Mis07,Mas12} and natural wastes \cite{Moh11} for
water treatment \cite{Nci09}, biosorption and biodegradation \cite{Pan11}
food contamination \cite{Sin2012} as well as medical applications such as
the chemical immobilization of bacteriophages on surfaces \cite%
{Nai12,Sin2013}.

The present paper is a extension of some of the ideas developed in our
previous works. We will take advantage of the recent progress in the
statistical theory of complex and deterministic chaotic systems as well as
the non equilibrium theory of Tsallis and others \cite{Tsa88,Tsa09}. We will
show that many of the isotherms used in the literature, especially in the
treatment of water, form a subfamily of the $Burr_{XII}$ distribution. This
will lead us to propose a generalization ($GBS$) \ of the $BS$ isotherm
replacing the exponential in the Weibull function by an deformed exponential
used now in the formulation of the nonextensive thermodynamics and other
complex systems theories. The same technique has helped us to elucidate the
universality of relaxation in disordered systems \cite{Bro04,Bro05}.and
formulate a fractional-time kinetics for $n$-order reaction systems \cite%
{Bro06}. As we will show, many of the isotherms used in the literature can
be obtained by giving well defined values to the parameters of this unified
isotherm. We will finally suggest that this $GBS$ isotherm has the same
degree of universality and the same stochastic and statistical foundation as
the Weron formula in the theory of relaxation in disordered systems \cite%
{Wer04}.

\section{The $Burr_{XII}$ distribution function.}

If we view the isotherm as a cumulative distribution function we can write
the isotherms in the following forms:%
\begin{equation}
\ \widetilde{\Theta }(P(C))=\frac{\Theta (P(C))}{\Theta _{\max }}%
=\int_{0}^{P(C)}\theta (p(c))dp(dc)
\end{equation}%
In eq.(1), $\theta (p(c))$\ is the relative sorbed quantity as the pressure
or concentration are increased in the gas or liquid phase in appropriate
units.\ The quantity $\Theta _{\max }$ is the maximum sorption capacity in
appropriate units. The $p(c)$ are supposed to be related thermodynamically
to a sorption energy variable $e:$\ 
\begin{equation}
p(c)\propto exp(-e/RT)
\end{equation}%
In an heterogeneous system, as we increase the pressure or the
concentration, the most active sites with the highest sorption energy are
first occupied until complete saturation. With a change of variable, one can
write%
\begin{equation}
\ \widetilde{\Theta }(P)=\int_{\Delta (P)}\theta (e)de
\end{equation}%
where $\Delta (P)$ is the range of energies involved at pressure $P$ and $%
\theta (e)$ is an energy dependent properly normalized distribution function$%
.$ This second formulation (eq.3) has been used to determine an empirical
formula for the sorption energy distribution \cite{Cer74,Jar86,Yag07,Kum11}.
In the following the variables $p$\ or $c$\ will be denoted by the greek
letter $\varkappa .$

We will now demonstrate that if we choose \ for $\tilde{\Theta}(\varkappa ),$%
\ the $Burr_{{\large XII}}$\ cumulative distribution function (cdf) many of
the physically sound isotherms used in the literature to generalize the
Freundlich formula can be recovered and a new generalized isotherm can be
proposed as a synthesis of the efforts of a few generations.

In probability theory and statistical sciences, the $Burr_{{\large XII}}$\
distribution is a continuous probability distribution for a non-negative
random variable \cite{Bur42}. It is also known in econometrics as the
Singh-Maddala distribution \cite{Mad96} where it has been used as a
generalization of the Pareto distribution for the graduation over the whole
range of incomes and is used to measure the level of inequality.

The $Burr_{{\large XII}}$\ distribution is a a member of a system of
continuous cumulative distribution (cdf) functions \ introduced by I.Burr in
1942 \cite{Bur42}. It has the form: 
\begin{equation}
F_{B}(x,a,b,c)=1-[1+c(\frac{x}{b})^{a}]^{^{-1/c}}
\end{equation}%
where $a,b,c$ are positive parameters. Its normalized probability density
function (pdf) $f_{B}(x,a,b,c)$ is obtained from 
\begin{align}
dF_{B}(x,a,b,c)\ & =f_{B}(x,a,b,c)dx=\frac{a}{b}(\frac{x}{b})^{^{\alpha
-1}}[1+c(\frac{x}{b})^{^{a}}]\ ^{-1-1/c}dx \\
dF_{B}(x^{a},a,b,c)\ & =f_{B}(x^{a},a,b,c)dx^{a}=\ (\frac{1}{b})^{^{a}}[1+c(%
\frac{x}{b})^{^{a}}]\ ^{-1-1/c}dx^{a}
\end{align}%
In previous papers \cite{Bro04,Bro05}, we have shown how it could be derived
from the maximum entropy principle using a generalization of the
non-extensive Tsallis entropy with appropriate constraints.

The cumulative distribution functions belonging to the Burr family are
solution of the general differential equation%
\begin{equation}
\frac{dF(x)}{d(x)}=f(x)=g(x)F(x)(1-F(x))
\end{equation}%
where $F(x)$\ and $g(x)$\ are continuous functions defined in specific
domains. This differential equation describes a birth and death function
modulated by a $g(x)$\ function which applied to a particular problem
depends on the nature of the phenomena and the influence of the environment.
The first and most studied of these differential equations is the famous
Verhulst logistic equation introduced in 1845 \cite{Ver1845} to mimic and
calculate population dynamics. In that case, $g(x)=1$\ and its solution is%
\begin{equation}
f_{V}(x)=1/(1+\exp (-x))
\end{equation}%
In its discrete form it has been one of the first model of deterministic
chaos \cite{Aus05}.

For the $Burr_{{\large XII\ }}$cdf, the twelfth one in the family, the
function $g(x)=$\ $\tilde{g}_{B}(x)/x$ has the form of an hyperbolic type
function, the function, $\tilde{g}_{B}(x)$\ \ varying smoothly between $a$\
and $a/c$.%
\begin{align}
g_{B}(x)& ={\large \ }\tilde{g}_{B}(x)/x\text{ \ \ where \ }{\large \ }%
\tilde{g}_{B}(x)=\text{ }\frac{(a(x/b)^{a})}{%
(1+c(x/b)^{a})(1-(1+c(x/b)^{a})^{(-1/c)})} \\
\text{ \ \ \ \ \ }{\large \ }\tilde{g}_{_{B}}(x)& \rightarrow a\text{ \ when 
}x\rightarrow 0\text{ \ \ and \ }{\large \ }\tilde{g}_{_{B}}(x)\rightarrow
a/c\text{ \ when }x\rightarrow \infty  \notag
\end{align}%
The $Burr_{{\large XII}}$\ distribution function has become a reference
distribution in complex and non equilibrium systems as the exponential and
Gaussian distributions are the reference distributions in equilibrium and
non interacting systems. The "dialectic" form of its differential equation
shows that it could be useful to deal with phenomena like for instance
epidemic propagation, population evolution, kinetics of complex reactions,
economic evolution, pharmacokinetic, cancer remission and obviously
sorption-desorption. It has been used extensively these last years in a
variety of chaos, nonlinear and nonequilibrium problems in quasi all fields
of pure and applied sciences including natural phenomena, meteorology,
hydrology, earthquake, economy, sociology and medicine.

An other interesting feature of the $Burr_{{\large XII}}$\ distribution is
the existence of two power laws tails, one for $x\rightarrow 0$\ with
exponent $a$\ and one for $x\rightarrow \infty $\ with exponent$\ \mu =a/c$.
It has a limited number of finite moments depending on the value of $\mu .$\
When $0<\mu <1,$ it has a heavy tail\ and belongs to the basin of attraction
of the family of stable L\'{e}vy distributions. It is to say, it has some
peculiar properties which have interesting consequences. L\'{e}vy functions
do not obey the traditional central limit theorem and an expectation value
of $x$\ cannot be defined. For higher values of $\mu $\ the average value
increases with the number of observations following a well defined power law 
\cite{Sor04}.

\section{The subfamily of the $Burr_{XII}$ distribution and the associated
isotherms.}

The $Burr_{XII}$ function (eq. 4) can generate a sub-family of cdf
distributions if one gives particular values to the two parameters $a$\ and $%
c$\ in $F_{B}(x,a,b,c)$. The case $a=$ $1$ and \ $c=0$ is simply the
exponential function. Some of these functions coincide with the form of well
known empirical isotherms: \ \ \ \ \ \ \ \ \ \ \ \ \ \ \ \ \ \ \ \ \ \ \ \ \
\ \ \ \ \ \ \ \ \ \ \ \ \ \ \ \ \ \ \ \ \ \ \ \ \ \ \ \ \ \ \ \ \ \ \ \ \ \
\ \ \ \ -for $c=0\ \ $%
\begin{eqnarray}
F_{B}(x,a,b,0) &=&F_{W}(x,a,b)=1-Exp(-\ (\frac{x}{b})^{a}) \\
\tilde{g}_{_{W}}(x &\rightarrow &0)\rightarrow a\text{ \ \ and \ }{\large \ }%
\tilde{g}_{W}(x\rightarrow \infty \text{ })\rightarrow \infty \text{ }
\end{eqnarray}%
This is a Weibull distribution.\ The corresponding isotherm in the sorption
literature is\ known as the $Brouers-Sotolongo$\ ($BS$) isotherm :%
\begin{equation}
\widetilde{\Theta }_{BS}({\Huge \varkappa })=[1-Exp(-(\frac{{\Huge \varkappa 
}}{b})^{a})]
\end{equation}%
If $\varkappa <<b$, one gets the $Freundlich$\ isotherm$\ $%
\begin{equation}
\Theta _{F}({\Huge \varkappa })=K_{F}\text{ }{\Huge \varkappa }^{a}
\end{equation}%
If moreover one puts $a=1$\ in eq.(12) , one gets the $Jovanovic$\ isotherm 
\cite{Jov69}%
\begin{equation}
\widetilde{\Theta }_{J}({\Huge \varkappa })=\ [1-Exp(-(\frac{{\Huge %
\varkappa }}{b}))]
\end{equation}%
-For $c=1$, one has:%
\begin{eqnarray}
F_{B}(x,a,b,1) &=&F_{HS}(x,a,b)=\frac{(\frac{x}{b})^{a}}{1+(\frac{x}{b})^{a}}%
=1-[1+(\frac{x}{b})^{a}]^{-1}\  \\
f_{B}(x,a,b,1) &=&f_{HS}(x,a,b)=\frac{a}{x}\frac{(\frac{x}{b})^{a}}{\ (1+\ (%
\frac{x}{b})^{a})^{2}}\ \ ;\ \ \ g_{HS}(x,a,b)=ax^{-1}
\end{eqnarray}%
which is called in probability theory the $loglogistic$ function. The
corresponding isotherms are the $Hill$\ , the $Langmuir-Freundlich$\ \ and $%
Sips$\ isotherms 
\begin{equation}
\ \widetilde{\Theta }_{LF}({\Huge \varkappa })=\ \frac{(\frac{{\Huge %
\varkappa }}{b})^{a}}{1+(\frac{{\Huge \varkappa }}{b})^{a}}=1-[1+(\frac{%
{\Huge \varkappa }}{b})^{a}]^{^{-1}}
\end{equation}%
-If both $a$\ and $c$\ are equal to 1: 
\begin{equation}
F_{B}(x,1,b,1)=F_{HS}(x,1,b)=F_{L}(x,b)=\frac{x\ }{b+x}\ =1-[1+(\frac{x}{b}%
)]^{^{\ -1}}
\end{equation}%
the corresponding isotherm is the $Langmuir$\ isotherm.%
\begin{equation}
\widetilde{\Theta }_{L}({\Huge \varkappa }){\large =\ }\frac{{\Huge %
\varkappa }{\Large \ }}{b+{\Huge \varkappa }}
\end{equation}%
As discussed in \cite{Bro05b}, the exponent $a$\ is related to the width and
shape of the sorption energy distribution which itself depends on the
heterogeneity of the substrate. In section 10 we will show that it defines
an effective temperature $T^{\ast }=T/a.$

In the isotherms we have just reviewed, the exponent $a$\ \ is supposed to
be constant and do not change with the evolution of the sorbed quantity $%
\Theta .$ This is a restrictive assumption. An isotherm derived from the
full $Burr_{{\large XII}}$\ would allow the characteristic exponent to vary
slowly from $a$\ to $a/c$.\ Therefore quite naturally a more realistic
isotherm based on the full $Burr_{{\large XII}}$\ distribution can be
proposed:%
\begin{equation}
\widetilde{\Theta }_{GBS}({\Huge \varkappa }){\Large \ =1-[1+c(}\frac{{\Huge %
\varkappa }}{b}{\Large )}^{a}{\Large ]}^{-1/c}
\end{equation}%
This generalized $BS$ isotherm has a unified character since it contains the 
$Langmuir$, the $Freundlich-Langmuir$, the $Hill$\ and the $Sips$\ \
isotherm \ and as we will see in the next section, the $Generalized$ $\
Freundlich-Langmuir$\ and the $Toth$ isotherms. The $GBS$\ isotherm can be
written in a more compact form 
\begin{equation}
{\Large \tilde{\theta}}_{GBS}\ ({\Huge \varkappa })=\{1-[1+c(\frac{{\Huge %
\varkappa }}{b})^{a}]^{-1/c}\}=1-\exp _{c}(-(\frac{{\Huge \varkappa }}{b}%
)^{a})
\end{equation}%
We have used the definition of the deformed exponential function introduced
in mathematics in the XIX century and appearing to day in the theory of many
complex systems 
\begin{eqnarray}
\exp _{c}(x) &=&(1-c\text{ }x)^{-\frac{1}{c}}\text{ \ \ if \ \ }1-c\text{ }x%
\text{\ }>0,\ \ 0\text{ otherwise \ } \\
\log _{c}(x) &=&\frac{1-x^{-c}\ }{c}\ \ \text{with }\exp _{c}(\ \log
_{c}(x))\ =\log _{c}(\ \exp _{c}(x))=x
\end{eqnarray}%
When $c=0$, one recovers the usual exponential. In the nonequilibriun
thermodynamic literature $c$\ = $q-1$\ where $q$\ is the Tsallis
nonextensive (nonadditive) entropy index. In the complex reaction
literature, $c$\ = $n-1$, where $n$\ is the effective fractional reaction
order.\ In the extreme value theory $c=\xi $\ , the shape parameter of the
distribution. We recover the $BS$\ isotherm $\theta _{BS}$\ if $c=0.$\ 

This new isotherm has four parameters $\theta _{\max },a,b,$and $c$\ which
have simple physical interpretation: $\theta _{\max }$is the maximum \
saturation sorbed quantity, $a$\ is the $Freundlich$ exponent which is
related to the width and shape of the sorption energy distribution, $c$\ is
related to the Tsallis entropic index $q=c+1$\ and when $q\neq 1$ is a
measure of the nonextensive character of the system. The coefficient $b$\ is
a $T$ dependent scale parameter and combined with $a$\ and $c$\ allows the
calculation of all the quantities characterizing the statistical
distribution: expectation, variance and moments, median, quantiles and some
other coefficients which measure quantitatively the way the sorption depends
on the concentration or the pressure. These useful expressions for the
analysis of isotherms are derived in the appendix. The value $a=1$ seperates
the distributions defining the isotherms in two groups.\ For $a\leq 1$ and
this includes the $Langmuir$ isotherm, the pdf is $L$-shaped while for $%
a\succ 1$, it is unimodular.\ This has a strong influence on the nature of
the sorption. We will show also in the appendix that the coefficient $b$ is
directly related to specific moment of the probability distribution. Finally 
$(a/c)$\ $=\mu .$ When $\mu $\ \ is $<1$\ it is the heavy tail (L\'{e}vy)
exponent which controls the upper behavior of the sorption curve and can
give rise to the particular properties of the functions belonging to the
basin of attraction of L\'{e}vy distributions \cite{Sor04}.

\section{An alternative differential equation for the $Burr_{XII}$\
function.\ }

It is easy to show using the definition of $F_{B}(x)$\ and $f_{B}(x)$\ that
the $BurrXII$ solution can be written%
\begin{align}
\frac{dF_{B}(x)}{dx}& =f_{B}(x)=\frac{a}{b}(\frac{x}{b}%
)^{^{a-1}}(1-F_{B}(x))^{^{c+1}\ }\  \\
\text{or \ \ \ }\frac{d\tilde{F}_{B}(x)}{dx^{a}}\ & =-\frac{1}{b^{a}}\tilde{F%
}_{B}(x)^{^{c+1}\ }\text{ \ \ \ }\tilde{F}_{B}(x)=1-F_{B}(x)
\end{align}%
Written in the time domain this differential equations is the stating point
of the \ fractal kinetic theory of ref. \cite{Bro06}.

-For $c=0$, we recover the Weibull equation which for $a=1$\ reduces to the
exponential form

-For $c=0$\ and $a=1,$\ we have the Tsallis differential equation used in
other physical problems \cite{Ped08} .%
\begin{equation}
\frac{d\tilde{F}_{T}({\Large \varkappa })}{dx}{\large =-}\frac{1}{b}{\large (%
}\tilde{F}_{T}{\large ({\Large \varkappa }))}^{q}{\large \ \ ,\ }\text{\ \ \ 
}{\large q=c+1}
\end{equation}%
The loglogistic form will be obtained when $c=1.\ $In that way the $%
Langmuir, $\ the $Freundlich-Langmuir$\ (or $Hill$\ or $Sips$), the $BS$\
and the $GBS$\ isotherms can be also derived from the solutions of the
alternative differential equation (23).

\section{The generalized Freundlich-Langmuir and Toth isotherms .}

A two exponents isotherm ($GFL$) generalizing the $Freundlich-Langmuir$ ($%
Hill$,$Sips$) isotherm was proposed by Marczewski and Jaroniec \cite{Mar83}.%
\begin{equation}
\tilde{\theta}_{GFL}({\Huge \varkappa })=(\frac{(\frac{{\Huge \varkappa }}{b}%
)^{n}}{1+(\frac{{\Huge \varkappa }}{b})^{n}})^{^{\frac{m}{n}}\text{\ }}\text{%
\ }
\end{equation}%
The corresponding $cdf$ function 
\begin{equation}
F_{G}(x)=(\frac{(\frac{x}{b})^{n}}{1+(\frac{x}{b})^{n}})^{^{\frac{m}{n}}%
\text{\ }}=(1+(\frac{x}{b})^{-n}\ )^{-\frac{m}{n}}\text{\ }
\end{equation}%
has the characteristics of a cdf. $F_{G}(0)=0,$\ $F_{G}(\infty )=1$

It appears that $F_{G}(x)$\ has the form of a $Dagun$\ function \cite{Dag77}
used concurrently with the $Burr_{{\large XII}}$\ equation in econometrics.\
It can be related to the $Burr_{{\large XII}}$\ function by a simple change
of variables. This will allow us to relate the isotherms obtained from the $%
GFL$\ isotherm form (26) to the ones already derived.

As $F_{B}(x)$, $F_{G}(x)$\ is also the solution of a first order
differential equation.\ Indeed one has 
\begin{equation}
\frac{dF_{G}(x)}{dx}=\frac{mb(\frac{(\frac{x}{b})^{n}}{1+(\frac{x}{b})^{n}}%
)^{^{\frac{m}{n}}\text{\ }}}{x(1+(\frac{x}{b})^{n})}=\frac{mb}{x}(\frac{x}{b}%
)^{^{-n}}F_{G}(x)^{^{\frac{n}{m}+1}}
\end{equation}%
We have moreover :%
\begin{equation}
\text{for }x\rightarrow 0,\text{ }\frac{dF_{G}(x)}{dx}\rightarrow mb(\frac{x%
}{b})^{m-1}\text{\ \ and \ for }x\rightarrow \infty ,\text{ }\frac{dF_{G}(x)%
}{dx}\ \rightarrow mb(\frac{x}{b})^{-n-1}
\end{equation}%
These asymptotic behaviors which are supposed to be the same as the ones of $%
\frac{dF_{B}(x)}{dx}\ $gives the relations between the exponents of the two
formulations.%
\begin{equation}
m=a\text{\ \ , \ }m/n=c\text{ \ , \ }n=a/c=\mu
\end{equation}

Starting from the $GLF.\ $isotherm equation, one can recover some of the
empirical isotherms : for $m=n=1,$\ the $Langmuir$\ isotherm, for $m=n$, the 
$Langmuir-Freundlich$\ or $Hill$\ isotherm. For $n=1$, the $Sips$\ isotherm
and for $m=1$\ the $Toth$ \cite{Tot95} isotherm.\ The first belongs to the
subfamily of the $Burr_{XII}$\ subfamily isotherms and have been already
considered.\ We will see now how the last one is linked to the $Burr_{%
{\large XII}}$\ function using the relations between the two probability
functions.

\section{$Dagum$ distribution versus $Burr_{XII}$ distribution.}

\ The $Burr_{XII}$\ $cdf.$\ and $pdf$ functions ()\ can be written%
\begin{eqnarray}
F_{B}(x,a,\beta ,c) &=&\ 1-[1+(\frac{x}{\beta })^{a}]^{-1/c}\text{ \ \ \ \
with \ \ \ \ }\beta =b(c^{-\frac{1}{a}}\text{)} \\
f_{B}(x,a,\beta ,c) &=&\frac{a}{\beta c}(\frac{x}{\beta })^{^{\alpha
-1}}[1+\ (\frac{x}{\beta })^{^{a}}]\ ^{-1-1/c}
\end{eqnarray}%
If we make the change of variables $x$\ $\rightarrow 1/x$\ \ and $\beta $\ $%
\rightarrow 1/\beta $\ in $\ (),$\ we get the $Dagum$\ $cdf.$and $pdf.:$\ \
\ \ \ 
\begin{eqnarray}
F_{D}(x,a,\beta ,c) &=&\ [1+(\frac{x}{\beta })^{-a}]^{-1/c} \\
f_{D}(x,a,\beta ,c) &=&-\frac{a}{\beta c}(\frac{x}{\beta })^{^{-\alpha
-1}}[1+\ (\frac{x}{\beta })^{^{-a}}]\ ^{-1-1/c}
\end{eqnarray}%
Therefore one has the relation%
\begin{equation}
F_{B}(x,a,\beta ,c)=1-F_{D}(1/x,a,1/\beta ,c)\text{ }\ ;\text{\ \ }%
f_{B}(x,a,\beta ,c)=-f_{D}(1/x,a,1/\beta ,c)\text{ }(1/x^{2})
\end{equation}%
and$F_{D}(x,a,\beta ,c)$ $\ =1-F_{B}(1/x,a,1/\beta ,c)\ \ ;$\ \ $%
f_{D}(x,a,\beta ,c)$ $\ =-f_{B}(1/x,a,1/\beta ,c)x^{2}$

The relation between the $Generalized$ $Freundlich-Langmuir$\ function and
the Burr$_{{\large XII}}$\ function can be written using the previous
results: 
\begin{equation}
(\frac{\ \ (\frac{x}{\beta })^{n}}{1+(\frac{x}{\beta })^{n}})^{^{\frac{m}{n}}%
\text{\ }}=\text{ }(1+(\frac{\beta }{x})^{^{n}})^{^{^{-\frac{m}{n}%
}}}=F_{D}(x,n,\beta ,\frac{m}{n})=1-F_{B}(\frac{1}{x},n,\frac{1}{\beta },%
\frac{n}{m}),\text{ \ }\beta =b(\frac{m}{n})^{^{-\frac{1}{n}}}
\end{equation}%
This allows the $GLF$ isotherm and the $Toth$\ \cite{Tot95} isotherm as well
as the equivalent $Oswin$ isotherm \cite{Osw46} used in food industry to be
part of the $Burr_{{\large XII}}$\ isotherm family.

The others empirical isotherms\ \cite{Red59,Rad72}: correspond to couple of
values $m$\ and $n$ in the general form (eq.26) which give non physical
asymptotic behavior and therefore cannot be used over the whole range of
concentration or pressure They might give excellent fit over a limited range
of data, like the popular Redlich-Peterson isotherm \cite{Red59}, but cannot
give reliable information over the whole sorption process. In our opinion,
as a logical consequence of our work these isotherms should be discarded
since we dispose now, with the unified $GBS$\ form (eqs.20,21), of a two
exponent isotherm with a solid theoretical foundation.

We can now derive quite simply the shape of the sorption energy distribution
giving rise to the various isotherms we have just derived.

\section{Sorption Energy Distributions.}

As we already discussed in a previous publication, starting from the
thermodynamic relation%
\begin{equation}
\varkappa =\text{exp}(-e/RT)
\end{equation}%
and using the probability theory relation%
\begin{equation}
\left\vert f_{p}(\varkappa )d\varkappa \right\vert \ \ =\left\vert
f_{E}(e)dE\right\vert \ 
\end{equation}%
it is possible to calculate the sorption energy distribution corresponding
to each isotherm. As discussed later, this sorption energy $e$ is the energy
which governs the macroscopic thermodynamic properties of the system.\ It is
not the microscopic site energies resulting from the atomic and molecular
interactions.

In that way we have obtained the following results:

- for the proposed $GBS$. isotherm derived from the Burr$_{{\large XII}}$\
distribution function: 
\begin{equation}
\phi _{GBS}(e)=\frac{a}{RT}(b^{-a}\exp (-ae/RT))(1+c(b^{-a}\exp
(-ae/RT)))^{-1-1/c}
\end{equation}

The other distributions can be obtained easily: \ \ \ \ \ \ \ \ \ \ \ \ \ \
\ \ \ \ \ \ \ \ \ \ \ \ \ \ \ \ \ \ \ \ \ \ \ \ \ \ \ \ \ \ \ \ \ \ \ \ \ \
\ -for $c\rightarrow 0,$\ we have the distribution corresponding to the $BS.$%
\ isotherm%
\begin{equation}
\phi _{BS}(e)=\frac{a}{RT}(b^{-a}\exp (-ae/RT))\exp (b^{-a}\exp (-ae/RT))\ 
\end{equation}%
-for $c\rightarrow 1,\ $\ we have the distribution corresponding to the $%
Hill $-isotherm :%
\begin{equation}
\phi _{H}(e)=\frac{a}{RT}(b^{-a}\exp (-ae/RT))(1+(b^{-a}\exp (-ae/RT)))^{-2}
\end{equation}

It is worth noticing that the $BS$. distribution has the form of the $Gumbel$
\cite{Gum54,Col01} (maximum) extreme value probability distribution function 
\begin{eqnarray}
\phi _{G}(e) &=&\frac{1}{\beta }Exp(-(\ \frac{e-\mu }{\beta }))Exp(-Exp(%
\frac{e-\mu }{\beta })) \\
\Phi _{G}(E) &=&Exp(-Exp(-(\frac{E-\mu }{\beta }))) \\
\text{with \ \ \ \ \ }\beta &=&RT/a\text{ \ \ \ \ \ \ and \ \ \ \ \ \ }\mu
=-RT\log b
\end{eqnarray}

The standard deviation of this function is well known%
\begin{equation}
\beta \pi \text{/}\sqrt{6}=(kT/a)\pi \text{/}\sqrt{6}
\end{equation}%
confirming the conclusions of reference \cite{Bro05b} about the physical
signification of the exponent $a$.

The function \ $\phi _{GBS}(e)$\ corresponding to the new proposed $GBS$\
isotherm is one member of the family of generalized $Gumbel$ functions. 
\begin{equation}
\phi _{GBS}(e)=\frac{a}{RT}(b^{-a}\exp (-ae/RT))(1+c(b^{-a}\exp
(-ae/RT)))^{-1-1/c}
\end{equation}%
\begin{eqnarray}
\Phi _{GBS}(E) &=&Exp_{c}(-Exp(-(\frac{E-\mu }{\beta }))) \\
\text{with \ \ \ \ \ }\beta &=&RT/a\text{ \ \ \ \ \ \ and \ \ \ \ \ \ }\mu
=-RT\log b
\end{eqnarray}%
It is the symmetric of the $Fisher-Tippett$ \cite{Fis28,Col01} generalized
extreme value cumulative distribution 
\begin{equation}
F_{FT}(E)=Exp(-Exp_{c}(-(\frac{E-\mu }{\beta })))
\end{equation}%
It is worth noticing that this last $GEV$ function (eq.48) could have been
obtained by using the $BS$\ isotherm (eq.12) and the Tsallis $c-$%
thermodynamic\ expression%
\begin{equation}
\varkappa _{i}=Exp_{c}(-e_{i}/RT)\text{ \ = \ }(1-c(e_{i}/RT))^{-1/c}\text{
\ \ \ with \ \ \ \ }c=q-1\text{ \ \ }
\end{equation}

To be complete we have calculated the energy distributions corresponding to
the $Freundlich-Langmuir$ isotherm%
\begin{equation}
\phi _{FL}(e)=\ \frac{m}{RT}\frac{{\Large b}^{-m}\exp {\Large (-me/RT)}}{(1+%
{\Large b}^{-n}\exp {\Large (-ne/RT))}^{^{\frac{m}{n}+1}}}
\end{equation}%
If $m=n$\ \ \ ($Hill,Sips$)%
\begin{equation}
\text{ }\phi _{H}(e)=\ \frac{n}{RT}\frac{{\Large b}^{-n}\exp {\Large (-ne/RT)%
}}{(1+{\Large b}^{-n}\exp {\Large (-nE/RT))}^{^{2}}}
\end{equation}%
If $m=1$\ ($Toth,Oswin$)%
\begin{equation}
\text{ }\phi _{T}(e)=\ \frac{1}{RT}\frac{{\Large b}^{-1}\exp {\Large (-e/RT)}%
}{(1+{\Large b}^{-n}\exp {\Large (-ne/RT))}^{^{\frac{1}{n}+1}}}
\end{equation}%
If $\ n=1$\ \ ($Generalized-Freundlich$)%
\begin{equation}
f_{GF}(e)=\ \frac{m}{RT}\frac{{\Large b}^{-m}\exp {\Large (-me/RT)}}{(1+%
{\Large b}^{-1}\exp {\Large (-e/RT))}^{^{m+1}}}
\end{equation}%
Some of the these distributions have been obtained earlier by various
authors without reference to the probability theory and using the Cerofolini
condensation approximation method \cite{Cer74}. The eq.(44)\ \ was derived
in \cite{Jar86}, eq.(42,54)\ was derived in \cite{Yag07} and eq.(55)\ \ was
derived in \cite{Kum11}. They have been used to determine numerically
sorption energy distributions from isotherm data and investigate the
thermodynamic nature of the sorption from the measured isotherms. The
detailed calculations require assumptions on the range of sorption energie,
the integrals being performed from $E_{min}$ to $E_{max}$ with respect to a
reference energy $E_{0}$. As $a$ and $c$ tend to $1$, one recovers the
Langmuir isotherm, the model with a unique sorption energy.\ Indeed the
energy probability density (eq.42 with $a=1$) is the derivative of a Fermi
function and tends to a Heaviside function as $T$ tends to $0$.\ The
corresponding pressure or concentration density function has a horizontal
asymptote at the origin.\ Physically this means that on a homogeneous
surface the pressure range over which sorption takes place (from a few
percents to complete coverage) at finite temperature, will be only of one or
two order of magnitude,and be narrower as $T$ decreases (and $b$ in our
notations decreases), an observation already discussed by Roginskii \cite%
{You62}.

\section{A new fractal Michaelis-Menten equation:}

The Michaelis-Menten equation :%
\begin{equation}
\frac{v}{\nu _{m}}{\Large =\ }\frac{s}{K+s}{\Large \ =\ }\frac{s/b}{1+s/b}%
\text{ \ \ \ }{\Large ,}\text{ \ \ \ }{\Large b=K}^{-1}
\end{equation}%
has long been the standard reference for biochemical kinetics \cite{Mic13}
describing the reaction of the substrate $S$\ on a free enzyme\ to form a
product $P\ $. Here $\nu =dP/dt$\ is the rate of an enzymatic reaction, $\nu
_{m}$ represents the maximum rate achieved by the system at maximum
(saturation) substrate conctration and $b^{-1}$ is the substrate
concentration at which the reaction rate is half \ of $\nu _{m}$. It has the 
$Langmuir$\ form. Using the inverse of the $Langmuir$\ function we obtain
immediately : 
\begin{equation}
s=b(\frac{\frac{\nu }{\nu _{m}}}{1-\frac{\nu }{\nu _{m}}})
\end{equation}%
If the reaction is allosteric (cooperative) so that $n$\ molecules of $s$\
bind to $E$, the kinetics are described by the Hill equation \cite{Mon65} 
\begin{equation}
\frac{\nu }{\nu _{m}}=\ \frac{(s/b)^{n}}{1+(s/b)^{n}}\ 
\end{equation}%
which has the form of the $loglogistic$\ distribution function (eq.15).
Using the inverse $loglogistic$\ function, we have%
\begin{equation}
s=b(\frac{\ \frac{\nu }{\nu _{m}}\ }{1-(\frac{\nu }{\nu _{m}})})^{^{^{\
1/n}}}
\end{equation}%
Nowadays, the $Hill$ formula is best thought as "interacting" coefficient
reflecting the extent of cooperativity among multiple bindng sites.\cite%
{Wei97} To account for the observed fractal time dependent rate equation due
to the spatially constraints of the reactants on the microscopic and
mesoscopic level, Savageau\ \cite{Sav97} proposed empirically the following
formula . 
\begin{equation}
s=b(\frac{\ \frac{\nu }{\nu _{m}}\ }{(1-(\frac{\nu }{\nu _{m}}))^{m}}%
)^{^{^{\ 1/n}}}
\end{equation}%
Although it is commonly used, this empirical function does not provide a
simple analytic inverse probability function and has been shown to exhibit
some numerical difficulties \cite{Mal03}. In the spirit of the present work
and the fractal kinetic equation \cite{Bro06}, we propose the following form
which has a better theoretical basis :%
\begin{equation}
\frac{\nu }{\nu _{m}}=\ 1-[1+c(\frac{x}{b})^{a}]^{-1/c}\ =exp_{c}(-(\frac{x}{%
b})^{a})\text{ \ with \ }c=q-1
\end{equation}%
\ and using the inverse of the $Burr_{{\Large XII}}$\ function%
\begin{equation}
F_{B}{\Large (x)}^{-1}{\Large =b[}\frac{(1-x)^{-c}-1}{c}{\Large ]}^{^{\frac{1%
}{a}}}
\end{equation}%
\ 
\begin{equation}
s=b[\frac{1-(1-(\frac{\nu }{\nu _{m}}))^{1}/^{c}}{c(1-(\frac{\nu }{\nu _{m}}%
))^{1/c}}]^{^{1/a}}
\end{equation}%
when $c=1$, this equation reduces to he $Hill$ equation and when $c=a=1$\ to
the Michaelis-Menten equation. This formulation accounts for a time
dependent rate. The difference with the Savageau formula is that the
exponent $1/c$ appears in the denominator and gives a correct asymptotic
behavior.

\section{Conclusions}

In this paper we have shown that a generalized isotherm having the
analytical form of a $Burr_{XII}$ cdf is able to generate a whole family of
empirical isotherms used in the literature to represent the sorption data of
a great number of solid-gas and solid-liquid sorbate-sorbent couples. Due to
the fact that the $Burr_{XII}$ and associated functions are used extensively
in econometrics,\ there exists on the market efficient nonlinear fitting
computing programs and the use of the $GBS$\ isotherm should make obsolete
the comparison, often with questionable linear fitting, of experimental
isotherms with the various approximations of this more general unified
isotherm. Practically since the $GBS$\ isotherm interpolates nicely between
the $BS$ $(c=0)$ and the $Hill-Sips$ $(c=1)$ isotherm and since the two ($%
a,b $) parameters isotherms give generally a reasonably good fit, one can
first try both of them and then using these partial results improve the fit
with the three ($a,b,c$) parameters $GBS$.In the same spirit a formally
correct fractal Michaelis-Menten equation has been proposed to deal with
catalytic enzyme reactions

The statistical expressions given in the appendix allows a mathematically
well defined characterization of the data. Extensions of the$Burr_{XII}$
have been proposed with extra parameters. They belong to the $Generalized$ $%
Beta$ $2$ distribution family and are legitimate cumulative probability
functions \ \cite{MCD84}.\ Such an extension which might be of interest for
huge number of data are irrelevant in sorption problems due to the
relatively small number of experimental data.

Another important conclusion of this study is that the energy distributions
giving rise to the $BS$\ and $GBS$\ isotherms belong to the family of
extreme value distributions. This is in agreement with the stochastic theory
of K.Weron $et$ $al$. \cite{Wer97,Wer02} which was developed for relaxation
in disordered medias and whose relation with the Tsallis nonequilibrium
thermodynamic theory has been discussed in references \cite{Bro04,Wer04}.
What matters in highly heterogeneous media is not the detailed microscopic
interactions but the extreme value distribution of interaction energies of
dynamically highly correlated mesoscopic clusters (on surfaces, patches,
islands). The relation between the phenomenological laws and their
microscopic causes has to go through the spatio- temporal scaling properties
of these intermediate cooperative regions. This representation allows to
average together a large number of extreme probabilistic events to form a
predictable picture of the behavior of the entire system. As a consequence,
the observed tail exponents $a$ and $a/c$ and the analytic form of the
equations describing the macroscopic properties are related to the extreme
value cluster energy distributions.\ The parameter $a$ defined an effective
temperature $T^{\ast }=T/a$ \ and $c=q-1$ is related to the Reyni-Tsallis
entropy factor $q$. In catalysis, the appearance of an effective temperature 
$T^{\ast }$ has been traced to the conditions at which the substrate was
prepared and annealed.\ The active centers regarded as defects once in
thermal equilibrium at temperature $T^{\ast }$ are "frozen" by sudden
cooling (quenching)$_{{}}$\ \cite{Sch57,You62}. More generally an effective
temperature $T^{\ast }\neq T$ expresses the fact that, due to the
frustrations induced by the geometry and the interactions, the couple
sorbate-sorbent is not in thermal equilibrium at the experimental
temperature $T$.

Finally as a consequence of these remarks, one has to be conscious that the
energy distributions which can be obtained from the isotherm using the
formulas derived in this paper are not the true microscopic sorption energy
distributions. It is an illusion to assume that macroscopic data such as a
sorption isotherm might give precise and detailed informations on the
microscopic geometry and\ atomic and molecular sorption interaction energy
of a highly heterogeneous sorbent. As we already said, they represent a
"mesoscopic" extreme value energy distribution of correlated clusters.

Two last remarks have to be made on the range of applicability of the
results of this paper. One has to emphasize that it deals with one aspect of
sorption i.e. the generalization of the $Langmuir$ isotherm to highly
heterogeneous surfaces and solid-liquid interfaces and in some cases of
complex composition of sorbates and sorbent. It concerns in particular most
works done in water and air decontamination research with pure or treated
natural products.

The sorption of simple molecules on smooth surfaces and well defined rough
surfaces \cite{Sah97,Sah97b} does not necessarily necessitate an elaborate
treatment as used in this paper and the analysis of its isotherms can bring
some partial information on the microscopic properties of the surface. In
many more complex systems, other phenomena such as wetting, capillarity
condensation in pores \cite{Sah96}. as well as diffusion, volume
condensation and multi-reactions effects might have to be considered.\ In \
those cases, more specific isotherm formulas have to be used \cite{Dab01}.
One should also be conscious that the analysis of data with the $GBS\ $%
isotherm are relevant only when applied to complete sets of data until
saturation.

\section{Appendix.}

The statistical quantities of all isotherms deriving from the unified $GBS$\
isotherm will be obtained simply by giving the corresponding values to the
coefficient $a$\ and $c$ to the statistical quantities of that isotherm
viewed as a cdf.

We can determine if and when an inflexion point will occur. The second
derivative changes sign at. 
\begin{equation}
\varkappa =b(\frac{a-1}{a+c})^{\frac{1}{a}}\text{ \ \ \ }a>1\text{ }
\end{equation}%
At that point when $a>1$, there will be an inflexion point in the isotherm.

The expression for the $k$-th moment is 
\begin{equation}
\langle \varkappa ^{k}\rangle =\frac{b^{k}}{c^{k/a}}\frac{\Gamma (1+\frac{k}{%
a})\Gamma (\frac{1}{c}-\frac{k}{a})}{\Gamma (\frac{1}{c})}\text{ \ \ \ \ }k<%
\frac{a}{c}
\end{equation}%
where $\Gamma $ is the Gamma function:

From eq.65,one can calculate the expectation value $\ \langle \varkappa
\rangle $ \ and the variance $\ \langle \varkappa ^{2}\rangle $ -$\ (\
\langle \varkappa \rangle $ )$^{2}.$

The mode it is to say the value of the variable corresponding to the highest
value of the distribution is 
\begin{equation}
\frac{b}{c^{1/a}}(\frac{a-c}{\frac{a}{c}+1})\text{ \ \ \ \ \ \ for \ \ \ }a>1
\end{equation}%
The inverse of the $Burr_{XII}$ function (eq.62) allows us to know the
pressure (or the concentration) corresponding to a given percentage of the
sorbed quantity.

One can then calculate the quantile $\varkappa _{p_{_{\%}}\ }$ solution of $%
\tilde{\Theta}_{GBS}(\varkappa _{p_{_{\%}}})=p_{\%}$, where $p_{_{\%}}$ is
the percentage of the sorbed quantity ranging from 0 to 1 . We have
therefore using the expression of the inverse $Burr_{XII\ }$ function
(eq.62) $\ $ 
\begin{equation}
\varkappa _{p_{_{\%}}\ }=b[\frac{(1-p_{_{\%}})^{-c}-1}{c}]^{^{1/a}}
\end{equation}%
This is the value of $\varkappa $ corresponding to a given percentage $%
p_{_{\%}}$ of the sorbed quantity.  With any two quantities $p_{\%1\text{ }}$%
and $p_{\%2\text{ \ \ }}$we have%
\begin{equation}
\frac{\varkappa _{p_{_{_{\%1}}}}}{\varkappa _{p_{_{\%2}}}}=[\frac{%
(1-p_{_{\%1}})^{-c}-1}{(1-p_{_{\%2}})^{-c}-1}]^{^{1/a}}
\end{equation}%
In the case of symmetric quantiles $\ \varkappa _{p_{_{\%1}}}$ and $%
\varkappa _{p_{_{\%2}}}=1-\varkappa _{p_{_{\%1}}}$, \ we have 
\begin{equation}
(\frac{\varkappa _{p_{_{\%}\ }}}{\varkappa _{\ 1-p_{_{\%}}}})_{GBS}=[\frac{%
(1-p_{_{\%}})^{-c}-1}{p_{_{^{\%}}}^{-c}-1}]^{^{1/a}}
\end{equation}%
which can be a useful quantity characterizing the sorption. For the two
simpler isotherms the$BS$ isotherm and the $Hill$ isotherm we have:%
\begin{equation}
(\frac{\varkappa _{p_{_{\%}}}}{\varkappa _{\ 1-p_{_{\%}}}})_{BS}=[\frac{\log
(1-p_{_{\%}})}{\log p_{_{\%}}}]^{^{1/a}}
\end{equation}%
\begin{equation}
(\frac{\varkappa _{p_{_{\%}}}}{\varkappa _{\ 1-p_{_{\%}}}})_{Hill}=[\frac{%
p_{_{\%}}}{1-p_{_{\%}}}]^{^{1/a}}
\end{equation}%
The last one has been obtained by S.Goutelle $et$ $al$.\ \ \cite{Gou08}

What to do when $a\leq 1$ and an expectation value cannot be calculated?\ \
We will show that it is nevertheless possible to calculate finite
characteristic quantities which can characterize the distribution.

Starting from the expression of the $kth$ moment.(eq.65) and choosing the
value%
\begin{equation}
\text{\ }k=a(\frac{1-c}{c})
\end{equation}%
the expression (65) yields%
\begin{equation}
c/b^{a}=\gamma =\langle \lbrack \varkappa ^{a(\frac{1-c}{c})}]\rangle ^{-%
\frac{c}{1-c}}=\langle \rangle _{a,c}
\end{equation}%
\begin{equation}
F_{B}(x,a,b,c)=1-[1+\ \gamma ^{\ }x^{a}]^{^{-1/c}}=1-[1+\ \langle \rangle
_{a,c}x^{a}]^{^{-1/c}}\text{ }
\end{equation}%
\begin{equation}
f_{B}(x,a,b,c)=c^{-1}\langle \rangle _{a,c}(x)^{^{\alpha -1}}[1+\langle
\rangle _{a,c}x^{a}]\ ^{-1-1/c}
\end{equation}%
The $pdf$ $Burr_{XII}$ function is the normalized function which maximizes
the entropy when the exponents $a$ and $c$ and the scale factor $b$ are
known.

Calculating the limits $c->0$ and $c->1$ we can calculate the corresponding
expressions for the $BS$ and $Hill-Sips$ isotherms.

In the first case when $c->0$ and using the properties of the function Gamma
in eq. (65), one has for any positive value $a$.

\begin{equation}
\langle \varkappa \rangle _{BS}=b\frac{1}{a}\Gamma (\frac{1}{a})
\end{equation}%
\begin{equation}
\langle \varkappa ^{k}\rangle _{BS}=b^{k}\Gamma (\frac{a+k}{a})
\end{equation}%
\begin{equation}
b^{a}=\langle \varkappa ^{a}\rangle _{BS}\ \ \ \ a>0
\end{equation}%
For the second case one gets 
\begin{equation}
\langle \varkappa \rangle _{HS}=b\frac{\pi }{a}\ Csc(\frac{\pi }{a})\text{ \
\ }a>1
\end{equation}%
\begin{equation}
\langle \varkappa ^{k}\rangle _{HS}=b^{k}\frac{\pi k}{a}\ Csc(\frac{\pi k}{a}%
)\text{ \ \ }a>k
\end{equation}%
It is convenient to have a relation between valid for all positive values of 
$a$.\ The simplest constraint resulting from the knowledge of the exponents $%
a$ and $b$ determined by experiment.is given using eq.(68) and the
properties of the Gamma function:%
\begin{equation}
b^{a}=\frac{4}{\pi ^{2}}\langle \varkappa ^{\frac{a}{2}}\rangle
_{HS}^{^{2}}\ \ \ a>0
\end{equation}

All these results can be obtained directly by performing the corresponding
integrals.$\ $We did not make use of the technique of "escort" probabilities
used in the work of Tsallis and disciples.

\ \ \ \ \ \ \ \ \ \ \ \ \ \ \ \ \ \ \ \ \ \ \ \ \ \ \ \ \ \ \ \ \ \ \ \ \ \
\ \ \ \ \ \ \ \ \ \ \ \ \ \ \ \ \ \ \ \ \ \ \ \ \ \ \ \ \ \ \ \ \ \ \ \ \ \
\ \ \ \ \ \ \ \ \ \ \ \ \ \ .

\begin{ack}
I would like to thank the three persons who have inspired me during this
work, Jean Paul Pirard from Li\`{e}ge who introduced me to the field of
adsorption and porous materials, Oscar Sotolongo from Havana who taught me
the Tsallis entropy and the L\'{e}vy distributions and Karina Weron from
Wroslaw who opened my eyes to the peculiar stochastic character of the
navigation through the micro-, the meso- and the macro world. I thank my
colleague Cedric Gommes for a critical reading of the manuscript
\end{ack}

\end{document}